\documentclass[10pt]{iopart} 
\usepackage[utf8]{inputenc}
\usepackage[english]{babel}
\usepackage{graphicx}
\expandafter\let\csname equation*\endcsname\relax
\expandafter\let\csname endequation*\endcsname\relax
\usepackage{amsmath}
\usepackage[colorlinks=true,allcolors=blue]{hyperref}
\usepackage{cite}
\begin{document}

\title[Sub-exponential distributions and the big jump principle]{Rare events in stochastic processes with sub-exponential distributions and the big jump principle}

\author{Raffaella Burioni $^{1,2}$ and Alessandro Vezzani$^{3,1}$}
\address{$^1$ Dipartimento di Scienze Matematiche, Fisiche e Informatiche, Università degli Studi di Parma, Parco Area delle Scienze, 7/A 43124 Parma, Italy}
\address{$^2$ INFN, Gruppo Collegato di Parma, Parco Area delle Scienze 7/A, 43124 Parma, Italy}
\address{$^3$ IMEM-CNR, Parco Area delle Scienze, 37/A, 43124 Parma, Italy}
\eads{ \mailto{raffaella.burioni@unipr.it} and \mailto{alessandro.vezzani@unipr.it}}


\begin{abstract}
Rare events in stochastic processes with heavy-tailed distributions are controlled by the big jump principle, which states that a rare large fluctuation is produced by a
single event and not by an accumulation of coherent small deviations. The principle has been rigorously proved for sums of independent and identically distributed random variables and it has recently been extended to more complex stochastic processes involving  L\'evy distributions, such as L\'evy walks and the L\'evy-Lorentz gas, using an effective rate approach. We review the general rate formalism and we extend its applicability to continuous time random walks and to the Lorentz gas, both with stretched exponential distributions, further enlarging its applicability. We derive an analytic form for the probability density functions for rare events in the two models, which clarify specific properties of stretched exponentials.
\end{abstract}

\maketitle

\section{Introduction}

Rare events in heavy-tailed distributions are an essential component in the description of stochastic systems. They occur naturally in 
geoscience, biology, ecology and economics, where they play a major role in risk assessment \cite{Gumbel,Hollander,Holger,Vulp,Embrechts,Lucilla1}. 

An interesting aspect of the mechanism driving rare fluctuations  is described by the {\it single big jump principle}. 
According to the principle, rare events in stochastic processes with heavy-tailed distributions originate from a single large  {\it jump} and not by the accumulation of many small coherent contributions, all in the same direction, as in thin-tailed processes. In particular, this means that  sums of independent and identically distributed (IID) random variables drawn from a heavy-tailed distribution  can exceed a very large value when one of them exceed that value.  In this form, the principle has been rigorously proved for sums of IID random variables with subexponential distributions \cite{Chistyakov,Foss} and in the presence of specific correlations \cite{Geluk,Clusel1}. In the physical literature the same case of IID has also been described in terms of condensation in probability space \cite{Maj2,filias,corberi}. 

Recently, using results from extreme values statistics,  the principle has been applied to a trap model  with L\'evy distributed trapping times \cite{WVBB19} and to  a L\'evy walk \cite{VBB19},  a classical stochastic process where the particle displacement is the sum of steps taken at constant velocity with duration drawn from a L\'evy  distribution \cite{zaburdaev}. The principle has also been extended to more general correlated stochastic dynamical processes by means of a heuristic "rate" approach, verified by numerical simulations \cite{VBB19,VBB19b}. In particular, the heuristic approach for the big jump principle has been applied to derive the analytic form of the tail in a generalized  L\'evy  walk with acceleration and deceleration along  the steps \cite{Albers,Sokolov}, and in the L\'evy Lorentz gas \cite{Fleurov,levyrand}, that models the motion of particles along a set of fixed scatterers spaced on a line according to a L\'evy  distribution.  Notably, in all cases that involve L\'evy  distributions the single step dynamics introduces a further dynamical scaling length \cite{levyrand} which controls the properties of the rare events and leads to strong anomalous diffusion \cite{castiglione}. Moreover, due to the single jump contribution, in L\'evy motion rare fluctuations strongly depend on the single step dynamics \cite{VBB19b}, in contrast with the usual universal form displayed by the bulk of the distribution, that follows from central limit theorems.  

The  L\'evy walk and the  L\'evy Lorentz gas are two examples of correlations among variables that goes beyond IID,  induced respectively by the motion along the steps and by the quenched positions of the scatterers, and it is remarkable that the big jump principle still holds in these frameworks.
The class of sub-exponential distribution is however much wider and L\'evy  distributions are only a particular case. 
Here, we take a further step in the analysis of the powerful big jump principle beyond IID. We consider a space-time coupled continuous time random walk \cite{zaburdaev,ctrw} and a Lorentz gas \cite{Fleurov,levyrand} and we focus on the case where the stochastic variables of the two processes follow a stretched exponential distribution. 

Stretched exponentials are typical sub-exponential functions which, however, decay faster than any power-law. They have been observed in a wide range of phenomena \cite{sornette}, such as in astrophysics, geology, biology and economical data-sets, and they encode a much smoother behavior.  In particular, the Weibull distribution \cite{weibull} represents, depending on the value of the parameter $\alpha$, a  sub-exponential stretched exponential for $\alpha<1$, an exponential Poisson distribution for 
$\alpha=1$ or a fast decaying distribution for $\alpha>1$. 

We apply the same techniques developed in \cite{VBB19,VBB19b} to the case of the Weibull distributions and we derive the analytic form of the Probability Density Function (PDF) of rare events,
both in the continuous time random walk and in the Lorentz gas.
 Moreover, with extensive numerical simulations we show that in the long-time asymptotic regime the analytic predictions fit very well the behavior of rare events for $\alpha<1$ while for $\alpha>1$, as expected, the form of the distribution matches the usual large deviation analysis in terms of rate functions \cite{Vulp} .  

Our results put into evidence the peculiar features of rare events in the presence of stretched exponentials, with respect of the power-law case. Stretched exponentials naturally contain a further characteristic length,  that enters the rare fluctuations prediction next to the dynamical scaling length of the stochastic process. As a consequence, the probability of rare events displays a more complex form, that can however be accurately predicted using the big jump formalism. Then, the system does not display strong anomalous diffusion \cite{castiglione}, as all moments of the distribution follow a Gaussian behavior. This means that in this case the non-Gaussian rare events are negligible with respect the typical (bulk) behavior and they does not affect the moments of the distribution.

The paper is organized as follows: in the next section we introduce the constant velocity continuous time random walk (VCTRW) and the  Lorentz gas with scatterers spaced according to a Weibull distribution. In Section \ref{BJ} we recall the rate approach to the big jump principle and in the next two sections we apply the principle both to the VCTRW and to the Lorentz gas, comparing the predictions with extensive numerical simulations.
In Section \ref{Weib} we discuss our results and outline the effects of the stretched-exponential distribution. Finally we present our conclusions.

\section{The VCTRW and the Lorentz gas with Weibull distributions}

We consider a stochastic process with random variables following a
stretched exponential Weibull distribution
\begin{equation}
\lambda_{\alpha,\tilde x}(x)= \frac{\alpha x^{\alpha-1}}{\tilde x^\alpha}e^{-\left(\frac{x}{\tilde x}\right)^\alpha}. 
\label{lambdaL}
\end{equation}
The characteristic scale $\tilde x$ and the exponent $\alpha$ define the distribution
$\lambda_{\alpha,\tilde x}(x)$.
For $\alpha<1$, $\lambda_{\alpha,\tilde x}(x)$ is sub-exponential, for $\alpha=1$ it corresponds to the Poisson process, while for $\alpha>1$ it is a thin-tailed distribution, with a fast decay.

We focus on two classical models for stochastic motion. The VCTRW \cite{zaburdaev,ctrw} is a sequence of steps $i$ starting a times $T_i$ ($T_{i+1}>T_i$). The duration of a step $t_i$ is drawn from the PDF $\lambda_{\alpha,\tilde t}(t_i)$ 
so that the step $i$ occurs at time $T_i=\sum_{j=1}^{i-1} t_i $ ($T_1=0$).
During each step, the walker moves with constant velocity $v$ in a random direction. 
The dynamics of the walker in the time interval $[T_{i+1},T_i]$ is 
$ r(T)=r(T_i)+ v_i (T-T_i)$, with $r(T_i)$ the position of the walker at $T_i$, and the random velocity $v_i=\pm v$, is drawn at $T_i$ with probability $1/2$. We set the initial condition  $r(0)=0$. and we consider as an observable the distance of the walker $R(T)$ from the starting point: $R(T)=|r(T)|$. Here, we consider one dimensional motion but the model can be generalized to any dimension.  

In the Lorentz gas \cite{Fleurov,levyrand}, we consider a 1-dimensional system of scattering points $i$ placed at positions $R_i$. We place the first scatterer on the origin $R_1=0$ and the 
others at $R_i=\sum_{j=1}^{i-1} L_j$ where the distances $L_j$ are random variables following the distribution $\lambda_{\alpha,\tilde L}(L_i)$. In this case, a continuous time random walk \cite{ctrw} is defined as follows.
The walker starts from the origin at $T=0$ and it moves with constant velocity $v>0$. When the walker reaches one of the scatterers, it can be transmitted or reflected. This means that if $R(T')=R_i$, for $T'<T<T''$ the evolution is $ r(T)=R_i+ v' (T-T')$ with velocity $v'=\pm v$ drawn at $T'$ with probability $1/2$. $T''$ is the time of the next scattering event ($R(T'')=R_{i+1}$ if $v'=+v$ or
$R(T'')=R_{i-1}$ if $v'=-v$). At $T''$ the velocity of the walker is again updated to $v''=\pm v$ with equal probability, apart from the case $R(T'')=R_1=0$ where $v''=+v$, i.e. we impose reflecting boundary condition at the origin (see \cite{Ub} for a possible generalization of the model to higher dimensions). 

For both models we study the PDF $P(R,T)$ to measure $R$, the distance from the origin, at time $T$. The PDF is averaged both on the fluctuations of the trajectories (positive and negative velocities) and on $t_i$ and $L_i$. 

The big jump principle applies to systems where $P(R,T)$ can be split in two terms describing respectively 
the "bulk" of the distribution and the rare events for very large $R$. The latter, $B(R,T)$,  for sub-exponential distributions can be obtained using the big jump principle. 
So we have:
\begin{equation}
P(R,T)\sim
\begin{cases}
\ell^{-1}(T)f(R/\ell(T)) &\mbox{if }\ R < \ell(T)\kappa(T) \\
B(R,T) &\mbox{if }\ R> \ell(T) \kappa(T) \\
\end{cases}
\label{sal}
\end{equation}
where $\ell(T)$ is the characteristic length of the process and $\kappa(T)$ is a slowly growing function of $T$ (e.g. a logarithmic function).
$P(R,T)$ converges in probability at large $T$ to $\ell^{-1}(T)f(R/\ell(T))$, 
a function which is significantly different from zero only for $R < \ell(T) \kappa(T)$.
This means that $\int_0^{\infty}|P(R,T)-\ell^{-1}(T)f(R/\ell(T))| dR \to 0$ for $T\to \infty$.  Therefore, $B(R,T)$ vanishes in measure: $\int_{\ell(T) \kappa(T)}^\infty |B(R,T)| dR \to 0$ for $T\to \infty$. However, since $B(R,T)$ describes $P(R,T)$ for $R\gg \ell(T)$, it can be relevant for higher moments of the distribution $\langle R^q (T)\rangle=\int_0^\infty P(R,T)R^q dR$ ($q>0$), since:
\begin{equation}
\langle R^q (T)\rangle \sim  \int_0^{\ell(T)\kappa(T)} \ell^{-1}(T)f(R/\ell(T)) R^q dR + \int_{\ell(T)\kappa(T)}^\infty B(R,T) R^q dR
\label{rP}
\end{equation} 
and the first term can be subleading with respect to the second integral for large $q$ \cite{castiglione}.
 
Since all the moments of the Weibull distribution are finite, we expect the bulk behavior of the model to be described by a normal diffusion with a Gaussian scaling function, i.e. $\ell(T)\sim T^{1/2}$ and $f(R/\ell(T))\sim \exp(-a(R/\ell(T))^2)$. For the  L\'evy walk this is a well known results while for the L\'evy-Lorentz gas this has been recently proved in \cite{Bianchi,Magdziarza,Bianchi2,Artuso}.

\section{The big jump principle}
\label{BJ}

The big jump principle  provides a general tools to evaluate $B(R,T)$ in the presence of sub-exponentially distributed random variables. Let us discuss the principle for a general stochastic process,  in a setting which is suitable to deal both with the VCTRW and the Lorentz gas in the same theoretical framework. We notice that in both models there exits a sequence $x_1, x_2, \dots$ of independent random variables drawn from a heavy tailed distribution $\lambda(x)$.  In the VCTRW $x_i$ is the duration of the step ($t_i$) while in the Lorentz gas $x_i$ is the distance between scatterers ($L_i$). In VCTRW we associate naturally to each variable $t_i=x_i$ a drawn time $T_i$. In the the Lorentz gas, the draw time  $T_i$ is more subtle, as the walker is travelling through a set of fixed scatterers, but it can be defined as the moment  when the walker crosses for the first time the scatterer placed at $R_i$. Indeed up to the time of the crossing $T_i$, the motion is independent of that $L_i$. We require  therefore that the variables $x_i$ are drawn from $\lambda(x)$ at a time $T_i$ with $T_i<T_j$ if $i<j$. The draw time $T_i$ is itself a stochastic variable which can depend, according to the model, also on the draws occurring before $T_i$, i.e. on $x_1,\dots, x_{i-1}$. 
 In general the PDF $P(R,T)$ of measuring the quantity $R$ at time $T$ ($R>0$ for the sake of simplicity) can be expressed as:
\begin{equation}
P(R,T)=\int \prod_i dx_i \lambda(x_i)  {\cal F}(R|T,\{x_i\})
\label{Pgen}
\end{equation}
where ${\cal F}(R|T,\{x_i\})$ is the probability of measuring $R$ at time $T$ given the sequence of random variables $\{x_i\}$.  Since because of causality $T_{n+1}$ does not depend on the $x_j$ with $j>n$, 
we can define the probability $w_n(x_1,\dots,x_n,T)$ that given the sequence $x_1,\dots,x_n$, we have $T_n<T<T_{n+1}$. Moreover, as noticed above, the value $R$ at time $T$ does not depends on the draws $n$ occurred at times 
$T_n>T$, and therefore  we can  
introduce ${\cal F}_n(R|T,x_1, \dots,x_n)$, that is the PDF to measure $R$ at time $T$ given  $T_n<T<T_{n+1}$.
We have then:
\begin{equation}
P(R,T)= \sum_{n=1}^{\infty} \int \prod_{i=1}^{i<n} d x_i \lambda(x_i)
w_n(x_1,\dots,x_n,T) {\cal F}_n(R|T,x_1, \dots,x_n)
\label{Pgen_2}
\end{equation}
Now we can define $\langle N(T) \rangle$ the average number of draws up to time $T$, i.e.
\begin{equation}
\langle N(T) \rangle = \sum_{n=1}^{\infty} n \int \prod_{i=1}^{i<n} d x_i \lambda(x_i)
w_n(x_1,\dots,x_n,T) 
\label{Pgen_3}
\end{equation}

Let us recover the L\'evy walk model in this unusual formalism. In that case, $x_i$ corresponds to the step duration $t_i$, while the draw times $T_i$ and the distance $R$ have been defined in the previous sections. 
We obtain therefore $w_n(t_1,\dots,t_n,T)=\theta(T-\sum_{i=1}^{n-1} t_i) \theta(\sum_{i=1}^{n} t_i-T)$, i.e. the 
probability that $T_n<T<T_{n+1}$ is $1$ if $\sum_{i=1}^{n-1} t_i<T<\sum_{i=1}^{n} t_i$ and $0$ otherwise. Notice that this is an alternative way to include the so called backward recurrence time in the calculation
\cite{zaburdaev,ctrw}.
Moreover the function ${\cal F}_n(R|T,t_1, \dots, t_n)$ is:
\begin{equation}
{\cal F}_n(R|T,t_1, \dots, t_n) = \int \prod_{i=1}^n d v_i \frac{1}{2}(\delta(v_i-v)+\delta(v_i+v))\delta(R-|\sum_{i=1}^{n-1} v_i t_i +v_n(T-\sum_{i=1}^{n-1} t_i)|)
\end{equation}
i.e. the distance $R$ is $|\sum_{i=1}^{n-1} v_i t_i +v_n(T-\sum_{i=1}^{n-1} t_i)|$ where the velocities $v_i=\pm v$ are drawn with probability $1/2$.

As said above, in the Lorentz gas, the $x_i$ are the distance $L_i$ between subsequent scatterer and  the draw times $T_i$ are the times at which the walker crosses for the first time the scatterer in $R_i$
 \cite{VBB19}. In this case, the explicit form of $w_n(L_1,\dots,L_n,T)$ and ${\cal F}_n(R|T,L_1, \dots,L_n)$ cannot be expressed in a simple way. However the big jump principle applies as well.

Since the distribution $\lambda(x)$ does not depends on time, 
the probability to draw $x$ at time $T_w$ is $p_{{\rm tot}}(x,T_w)=  r_{N}(T_w)\cdot \lambda(x) $, where  $r_{N}(T_w)=d \langle N(T_w)\rangle d T_w$ is the draw rate. Then, the principle poses that as $R\gg \ell(T)$ the only important contribution to $B(R,T)$  is the biggest jump and therefore we neglect all the jumps occurring before and after that jump. We call ${\cal P}(R|T,x,T_w)$ the probability that a process, driven by the single jump $x$ starting at $T_w$, takes the walker in $R$ at time $T\geq T_w$.  As a result, $B(R,T)$ is determined as:
\begin{equation}
B(R,T) \sim \int dx \int_0^T dT_w p_{{\rm tot}}(x,T_w) {\cal P}(R|T,x,T_w)
\label{BigJump}
\end{equation}
Hence, $B(R,T)$ is evaluated by summing over all the paths ($x$ and $T_w$) that in a single jump  bring the process in $R$ at time $T$. These paths, described by ${\cal P}(R|T,x,T_w)$, can be very complex, as they include all the correlations and non-linearities of the model. However, since only one draw is non-vanishing, 
Eq. \eqref{BigJump} is much simpler than the original problem in Eq.s (\ref{Pgen},\ref{Pgen_2}) where a (possibly
infinite) series of stochastic variables have to be taken into account, so an analytic approach is often feasible.
In particular, the rate  $r_{N}(T_w)$ is the only information on the bulk dynamics that we retain to apply the principle, so in practice we disentangle the rare event from the rest of the dynamics.

In the following section we provide an expression for $p_{{\rm tot}}(x,T_w)$ and ${\cal P}(R|T,x,T_w)$ both for the VCTRW and the Lorentz gas and  we test our analytical prediction against extensive numerical simulations. Notice however that a general rigorous approach providing the shape of ${\cal P}(R|T,x,T_w)$ for a generic process satisfying Eq. \eqref{Pgen_2} is still lacking.

\section{The VCTRW with Weibull distribution and the big jump principle}
\label{GLW-BJ}

\begin{figure}
\centering
	\includegraphics[width=0.80\textwidth]{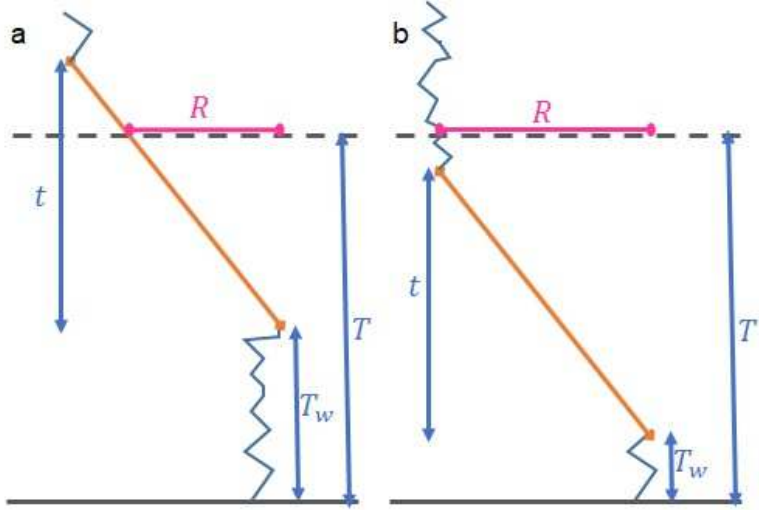}
	\caption{
		The big jump starts at time $T_w$ and it can either lead you to the
		time horizon of the L\'evy walk $t>(T-T_w)$ (panel a) or it may stop before the
		of the process $t<(T-T_w)$ (panel b). The magenta line represent the distance $R$ covered 
		in the big jump at time $T$. In the process in panel a 
    $R=v(T-T_w)$ while in panel b $R=vt$. }
	\label{LWfig}
\end{figure}

\begin{figure}
\centering
	\includegraphics[width=0.49\textwidth]{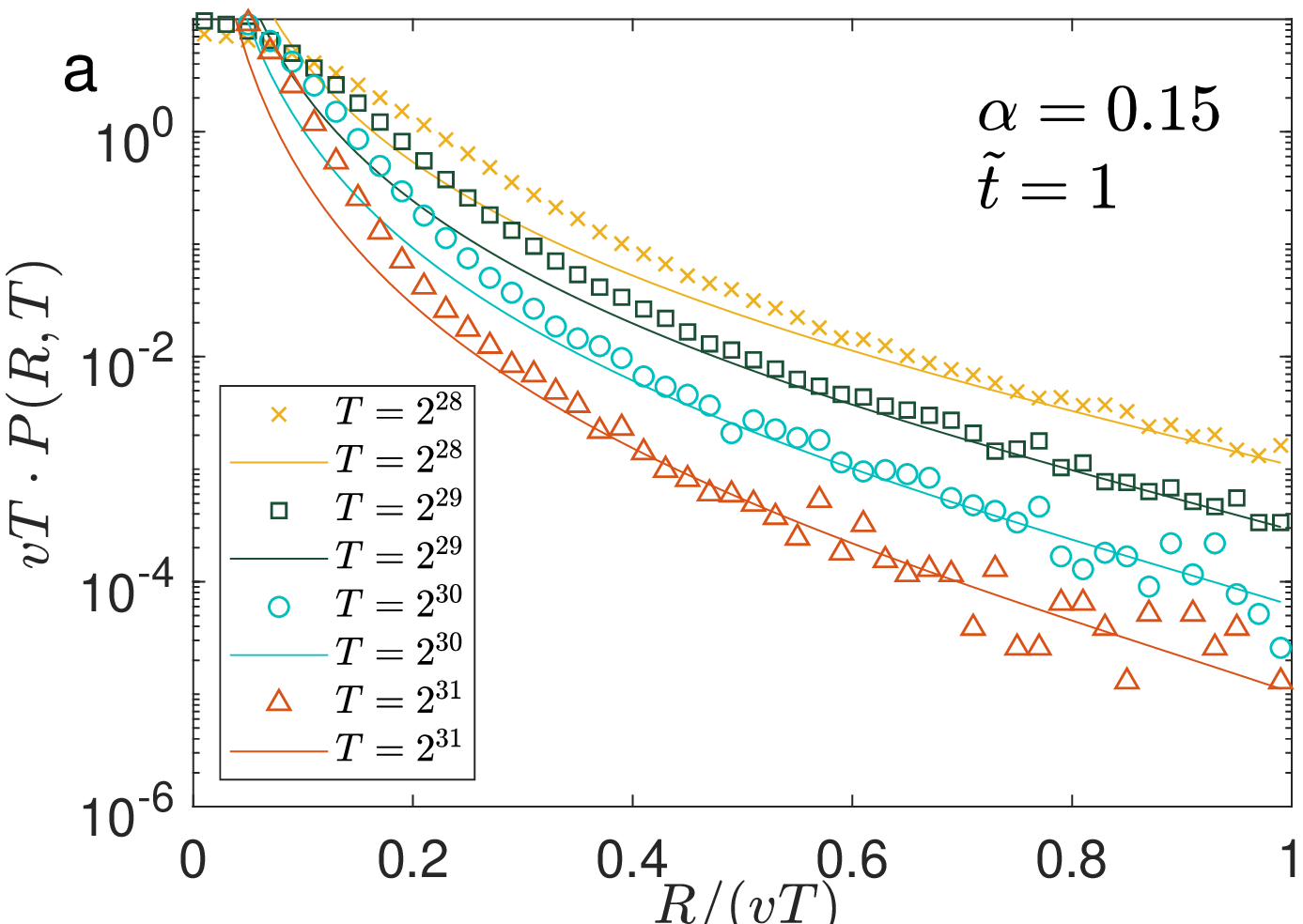}
	\includegraphics[width=0.49\textwidth]{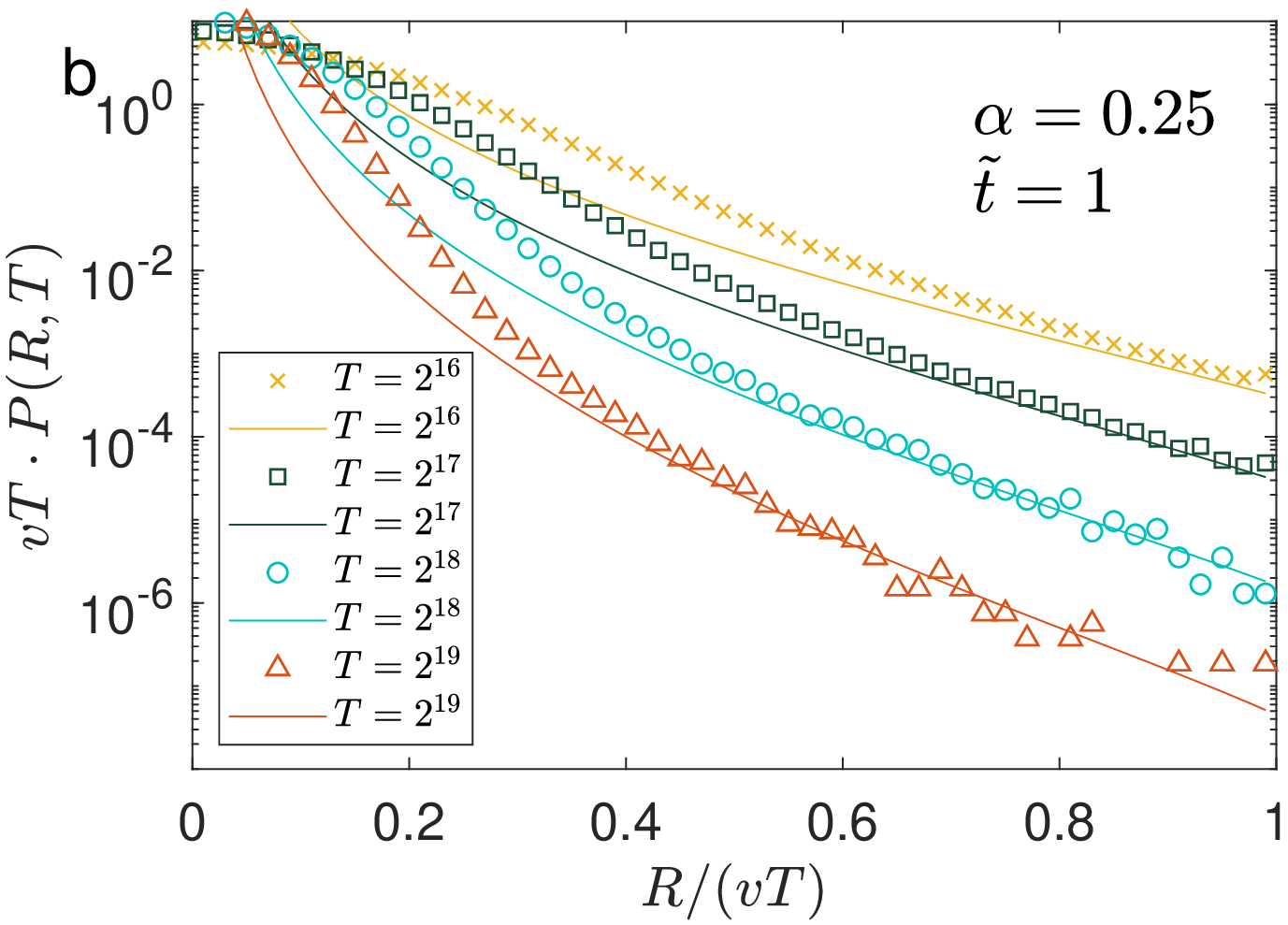}
	\caption{ The far tails of the distributions $P(R,T)$ with Weibull distribution of time durations and $v=1$, in particular $\alpha=.15$ with $\tilde t=1$ (panel a) and $\alpha=.25$ with $\tilde t=1$ (panel b). We plot the PDF as a function of $R/(vT)$ and we multiply the density by $vT$ in order to plot different dataset in the same plot. The thick continuous lines represent the estimate of the far tail obtained with the big jump approach: i.e. $vT G(R/(vT),\tilde t/T)$ described by formula (\ref{BJ1}). }
	\label{LW2fig}
\end{figure}

Let us derive the tail $B(R,T)$ by applying the big jump principle to the CRTW. We can neglect the motion of the walker before and after the single large contribution of duration $t$, since this is the only contribution to the displacement.
As show in Figure \ref{LWfig} two kind of processes can bring the walker in $R\gg\ell(T)$ at time $T$. In the first process, the step duration $t$ is larger than $T-T_w$ ($t>(T-T_w)$) so the walker is still moving along the big jump at $T$ and $R=v (T-T_w)$. In the second path in Fig. 1 $t<(T-T_w)$, the walker ends its motion at $t$ so that $R=v t$. 
Introducing the Kronecker $\delta$-function and the Heaviside $\theta$-function we get:
\begin{equation}
{\cal P}(R|T,t,T_w)  =  \delta(R-v(T-T_w)) \theta(t-(T-T_w))+ \delta(R-vt) \theta((T-T_w)-t)
\label{Lw1}
\end{equation}
where the first and the second terms correspond to the first and the second path respectively.
For the Weibull distribution in Eq. \eqref{lambdaL} the average duration of a step $\langle t \rangle=\int t \lambda_{\alpha,\tilde t}(t) dt$ is finite and the jump rate is constant, in particular $r_{N}(T_w) =\langle t\rangle^{-1}$. Plugging $p_{{\rm tot}}(t,T_w)=\lambda_{\alpha,\tilde t}(t) /\langle t\rangle$ and Eq. \eqref{Lw1} into formula \eqref{BigJump} we get:
\begin{equation} 
B(R,T)=\frac{e^{-\left(\frac{ R }{ v \tilde t }\right)^\alpha} }{v\langle t \rangle  }
\left(1+\alpha \left( \frac{ R }{ v \tilde t }\right)^\alpha \left(\frac{ vT }{ R }-1 \right)
\right)=F\left(\frac{ R }{ v T },\frac{ R }{ v \tilde t }\right)
\label{BJ1} 
\end{equation}
In Eq. \eqref{BJ1} two characteristic lengths are present: $vT$ and $v \tilde t$. In the usual power law L\'evy case, $B(R,T)$ is determined by the dynamic scaling length $vT$ only.  The Weibull distribution, indeed, contains a characteristic  length $v \tilde t$ which enters in the dynamical evolution at any distance $R$. We can rewrite Eq. \eqref{BJ1} as $F(\frac{ R }{ v T },\frac{ R }{ v \tilde t })=G(\frac{ R }{ v T },\frac{ \tilde t }{ T })$. In Fig. \ref{LW2fig} for $\alpha<1$, we compare the analytic prediction $G(\frac{ R }{ v T },\frac{ \tilde t }{ T })$ with a numerical Montecarlo estimate of the far tail of $P(R,T)$. In the long time limit, simulations fully agree with the big jump formalism.

\section{The  Lorentz gas and the big jump principle}
\label{LLG-BJ}

\begin{figure}
\centering
	\includegraphics[width=0.60\textwidth]{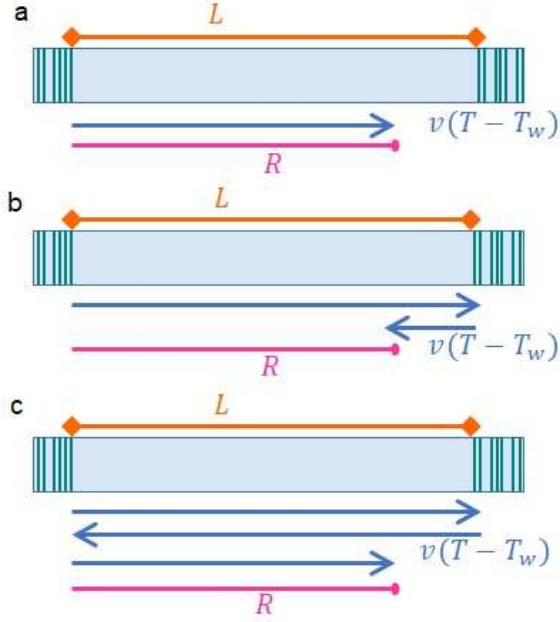}
	\caption{The single big jump process for in L\'evy Lorentz gas. Panel a,b and c refer to the cases with 0, 1 and 2 reflections respectively. The distance of the scattering points $L$ corresponding to the big jump is in orange, the final position $R$ is in magenta and the total traveled distance in the big jump $v(T-T_w)$ is in blue.}
	\label{LLGjump}
\end{figure}

Let us introduce the big jump approach for the Lorentz gas.
The independent random variables are the distances between consecutive scatterers $L_i$ and a jump occurs when the walker crosses a scatterer that has never been reached before. Hence $r_{N}(T_w)$ is the rate at which new scatterers are crossed by the walker. It has been shown in \cite{levyrand} that
for a distribution with finite average distance between sites $\langle L \rangle  = \int dL L \lambda_{\alpha,\tilde L}(L)<\infty$ we have at large $T_w$:
\begin{equation}
r_{N}(T_w) =
(T_w  \tau_0)^{-\frac{1}{2}}   
\label{ptw}
\end{equation}
where the time $\tau_0$ is a constant. Notice that here the rate depends on time.
The big jump approach states that $B(R,T)$ is determined by calculating the probability that the walker reaches a distance $R$ at time $T$ having crossed
at time $T_w$ for the first time a scattering point which is followed by a large jump of length $L\gg \ell(t)$. All the other distances between the scatterers are negligible with respect to $L$ and  therefore, up to time $T_w$, the motion of the walker can be ignored.  After crossing this long gap, the borders of the gap act as a perfect reflective walls on the walker. 
Indeed, for a recurrent random walk, the probability of not being reflected is vanishing. So, the motion, shown in Fig. \ref{LLGjump}, is the following:
up to time $T_w$ the walker remains at the starting point, then it bounces back and forth in the gap of length $L$ for a time $T-T_w$. The final position $R$ is denoted in magenta, the jump length $L$ in orange and the total distance covered by the walker $v(T-T_w)$ in blu. Using this physical insight on the motion of the walker we have $B(R,T)=\sum_{n=0}^\infty B_n(R,T)$ 
where $B_n(R,T)$ is the contribution of the path with $n$ reflections.

\begin{figure}
\centering
	\includegraphics[width=0.49\textwidth]{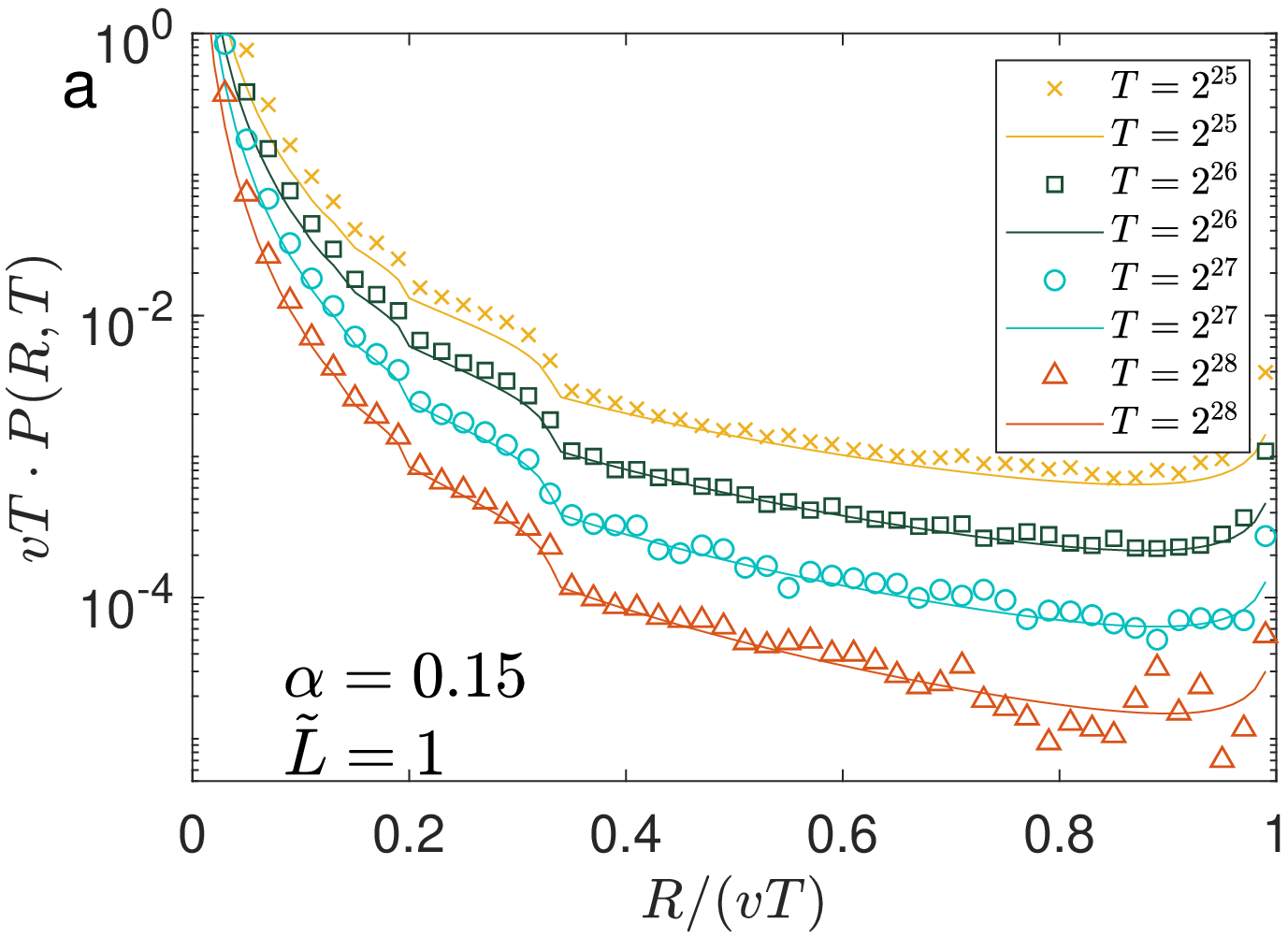}
	\includegraphics[width=0.49\textwidth]{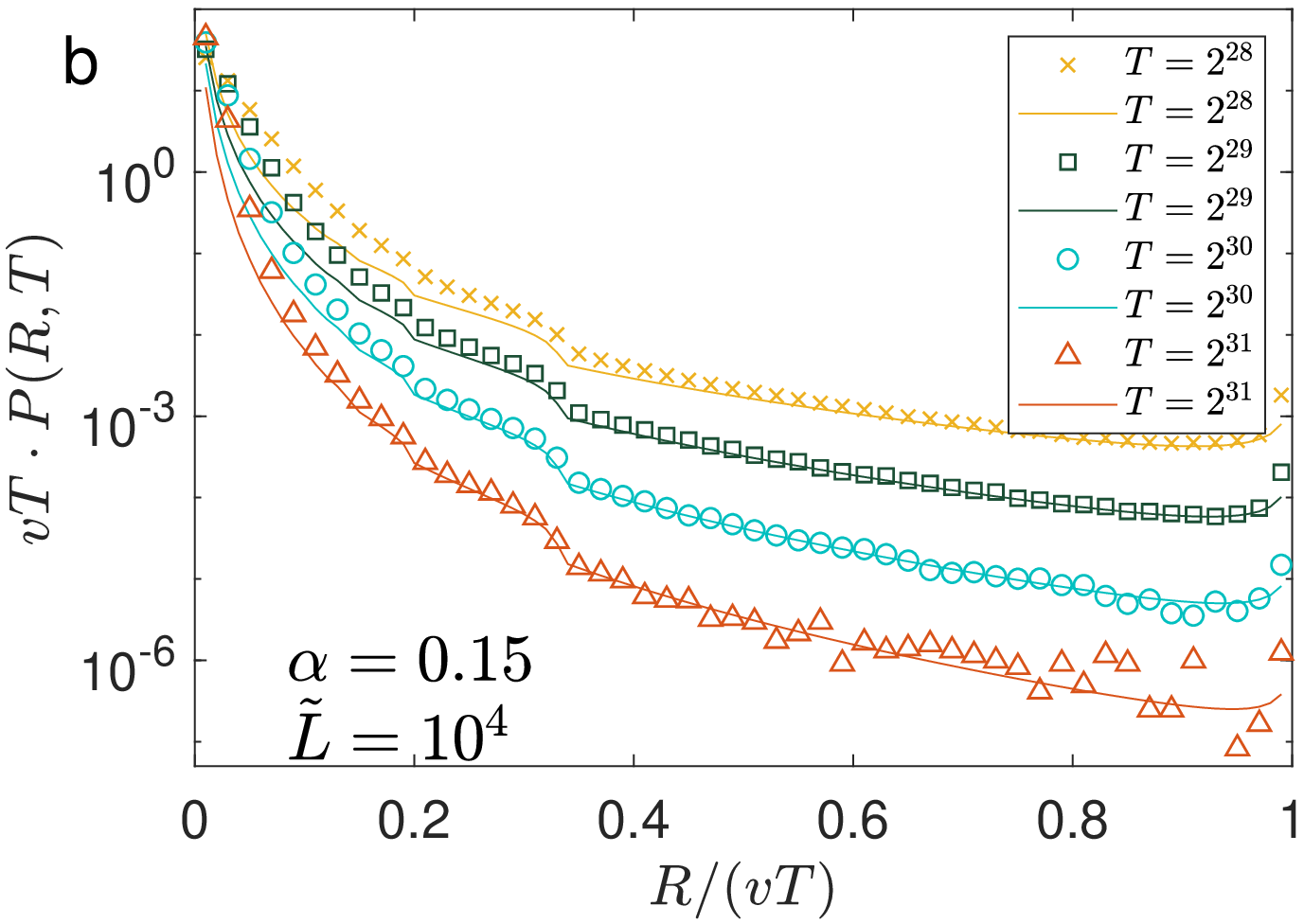}
	\caption{ The far tails of the distributions $P(R,T)$. We fix $v=1$ in panel a we consider a Weibull distribution with $\alpha=.15$ and $\tilde L=1$  and in panel b $\alpha=.25$ with $\tilde L=10^4$. We plot the PDF as a function of $R/(vT)$ and we multiply the density by $vT$ in order to plot different datasets in the same plot. The continuous lines represent the estimate of the far tail provided by the big jump approach: i.e. $vT G_\alpha(r/(vT),\tilde L/(vT))$ described by formula (\ref{single_j2}). }
	\label{LLfig}. 
\end{figure}

In particular panel a in Fig. \ref{LLGjump}  describes the case $v(T-T_w)<L$ where no reflection are present and $R=v(T-T_w)$. The contribution of this path to $B(R,T)$ is
\begin{equation}
\begin{aligned}
B_0(R,T) & = 
\int_0^T dT_w \int dL \lambda_{\alpha,\tilde L}(L) r_{N}(T_w)  \delta(R-v(T-T_w)) \theta(L-v(T-T_w))= \\
&=\int_R^\infty dL \lambda_{\alpha,\tilde L}(L) \frac{r_{N}(T-R/v)}{v}   \sim
\frac{1}{(\tilde L v \tau_0)^{{\frac{1}{2}}}} e^{-\left(\frac{R}{\tilde L}\right)^\alpha} \frac{ (\tilde L/(vT))^{\frac{1}{2}}}{ (1-R/(vT))^{{\frac{1}{2}}}} = \\
& =G_{0,\alpha}\left(\frac{R}{vT},\frac{\tilde L}{v T}\right)  
\label{gnpr0}
\end{aligned}
\end{equation}   
The number of reflections equals $n$ if $n L<v(T-T_w)<(n+1) L$ (see panels b and c for the case of $n=1$ and $n=2$) and
for $n$ odd $R+v(T-T_w)=(n+1)L$ so we obtain:
\begin{equation}
\begin{aligned}
B_n(R,T) & = 
\int_0^T dT_w \int dL \lambda_{\alpha,\tilde L}(L) r_{N}(T_w) \delta(R+v(T-T_w)-(n+1)L) \\
& \ \ \ \ \ \ \theta((n+1)L-v(T-T_w)) \theta(v(T-T_w)-nL)= \\
&=  \int_0^T  \frac{dT_w}{n+1} \lambda\left(\frac{R+v(T-T_w)}{n+1}\right) r_{N}(T_w)\theta(R)
\theta(v(T-T_w)-nR)=\\
&= \frac{1}{(\tilde L v \tau_0)^{1/2}} \frac{\alpha \theta\left(1- n\frac{R}{vT}\right)
\left(\frac{\tilde L}{vT}\right)^{1/2-\alpha}}{(n+1)^{\alpha}}\\
& \ \ \ \ \ \  \int_{0}^{1-n\frac{R}{vT}} e^{- \frac {\left( 1+\frac{R}{vT}-t_w\right)^\alpha}{\left((n+1)\frac{\tilde L}{vT}\right)^\alpha}} \frac{ dt_w}{t_w^{{\frac{1}{2}}} \left(1-t_w+\frac{R}{vT}\right)^{1-\alpha}}=\\
& =  G_{n,\alpha}\left(\frac{R}{vT},\frac{\tilde L}{vT}\right)
\end{aligned}
\label{gnpro}
\end{equation}  
where we introduce the integration variable $t_w=T_w/T$. We notice that  the $\theta$-function   $\theta(1- n\frac{R}{vT})$ means that the contribution is vanishing for $vT<nR$. On the other hand for $n$ even we have  $v(T-T_w)-R=nL$ and performing the analogous calculation we get:
\begin{equation}
\begin{aligned}
B_n(R,T) &=
 \frac{\alpha \theta\left(1- (n+1)\frac{R}{vT}\right)
\left(\frac{\tilde L}{vT}\right)^{1/2-\alpha}}{(\tilde L v \tau_0)^{1/2} n^{\alpha}}\\
& \ \ \ \ \ \   \int_{0}^{1-(n+1)\frac{R}{vT}} e^{- \frac {\left( 1+\frac{R}{vT}-t_w\right)^\alpha}{\left(n\frac{\tilde L}{vT}\right)^\alpha}} \frac{ dt_w}{t_w^{\frac{1}{2}} \left(1-t_w-\frac{R}{vT}\right)^{1-\alpha}}=\\
& = G_{n,\alpha}\left(\frac{R}{vT},\frac{\tilde L}{vT}\right)
\end{aligned}
\label{gnpre}
\end{equation}  
Summing the contributions in Eq.s (\ref{gnpr0}-\ref{gnpre}) we obtain:
\begin{equation}
B(R,T)=\sum_{n=0}^\infty B_n(R,T)=G_{\alpha}\left(\frac{R}{vT},\frac{\tilde L}{v T}\right).
\label{single_j2}
\end{equation}  
Again we notice that, as for VCTRW, $B(R,T)$ does not present a single scaling. Figure \ref{LLfig} compares numerical Montecarlo estimates of the far tail of $P(R,T)$ with the analytic prediction $G_\alpha(\frac{ R }{ v T },\frac{ \tilde L }{ (vT) })$, showing a nice agreement in the long time limit.
Fig.\ref{LLfig} shows the non analytic behavior predicted by the big jump approach when
$\frac{R}{vT}=\frac{1}{2m+1}$. These singularities have been observed also with power law distributions and they are produced by the reflections occurring in the big jump dynamics.

\section{The effects of the Weibull distribution}
\label{Weib}

The results of the previous sections put into evidence some relevant peculiarity of the big jump principle for a Weibull distribution. In this case, a simple scaling approach characterized by a single dynamical scaling length is not feasible.
Indeed, beside the dynamical scaling length $vT$ a second characteristic length is introduced in the system by the Weibull distribution. This is also confirmed by Figures \ref{LW2fig} and \ref{LLfig} which show that the shape of $P(R,T)$ explicitly depends on $T$ not only through a simple multiplicative prefactor. 

Another relevant feature is that in this case there are no anomalous moments. Both for the VCTRW and 
for the Lorentz gas when plugging the analytic expression for $B(R,T)$ into Eq. \eqref{rP}, the second integral $ \int_{\ell(T)\kappa(T)}^\infty B(R,T) R^q dR$ is exponentially sub-leading at large $T$
with respect to the first one $\int_0^{\ell(T)\kappa(T)} \ell^{-1}(T)f(R/\ell(T)) R^q dR$ so we obtain 
$\langle R^q(T) \rangle \sim \ell(T)^q = t^{q/2}$. Therefore, in the stretched exponential case we cannot apply  the techniques developed in \cite{Eli1,Eli2} which are based on the summation of the strongly anomalous moments, to calculate the far tail $B(R,T)$.
Finally, in this case the far tail is not described by an infinite density \cite{Eli1,Eli2}: indeed we have that the integral $\int_0^\infty B(R,T)dR$ is finite since for both models $B(R,T)\sim R^{1-\alpha}$ for $R\sim 0$.

\begin{figure}
\centering
	\includegraphics[width=0.49\textwidth]{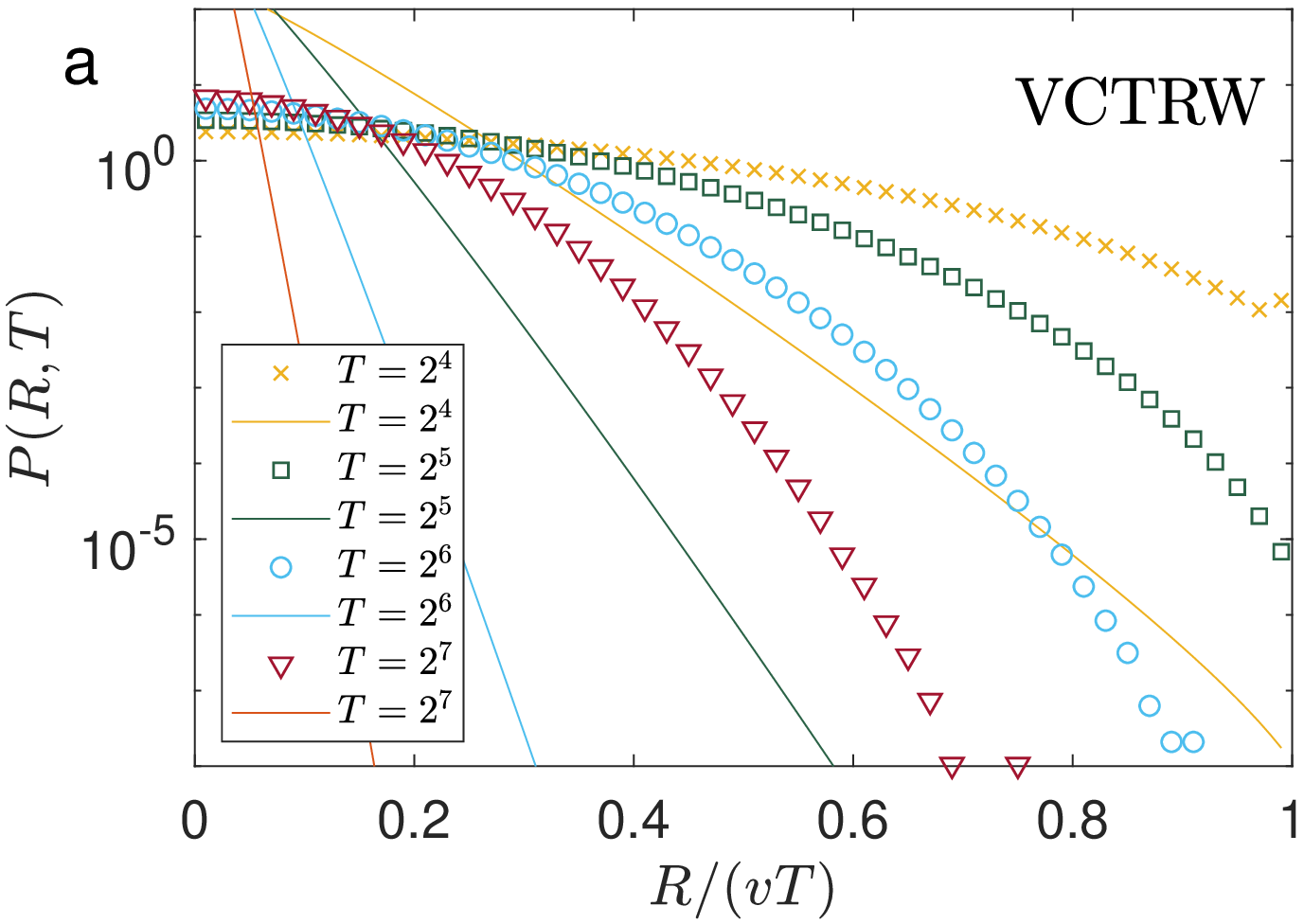}
	\includegraphics[width=0.49\textwidth]{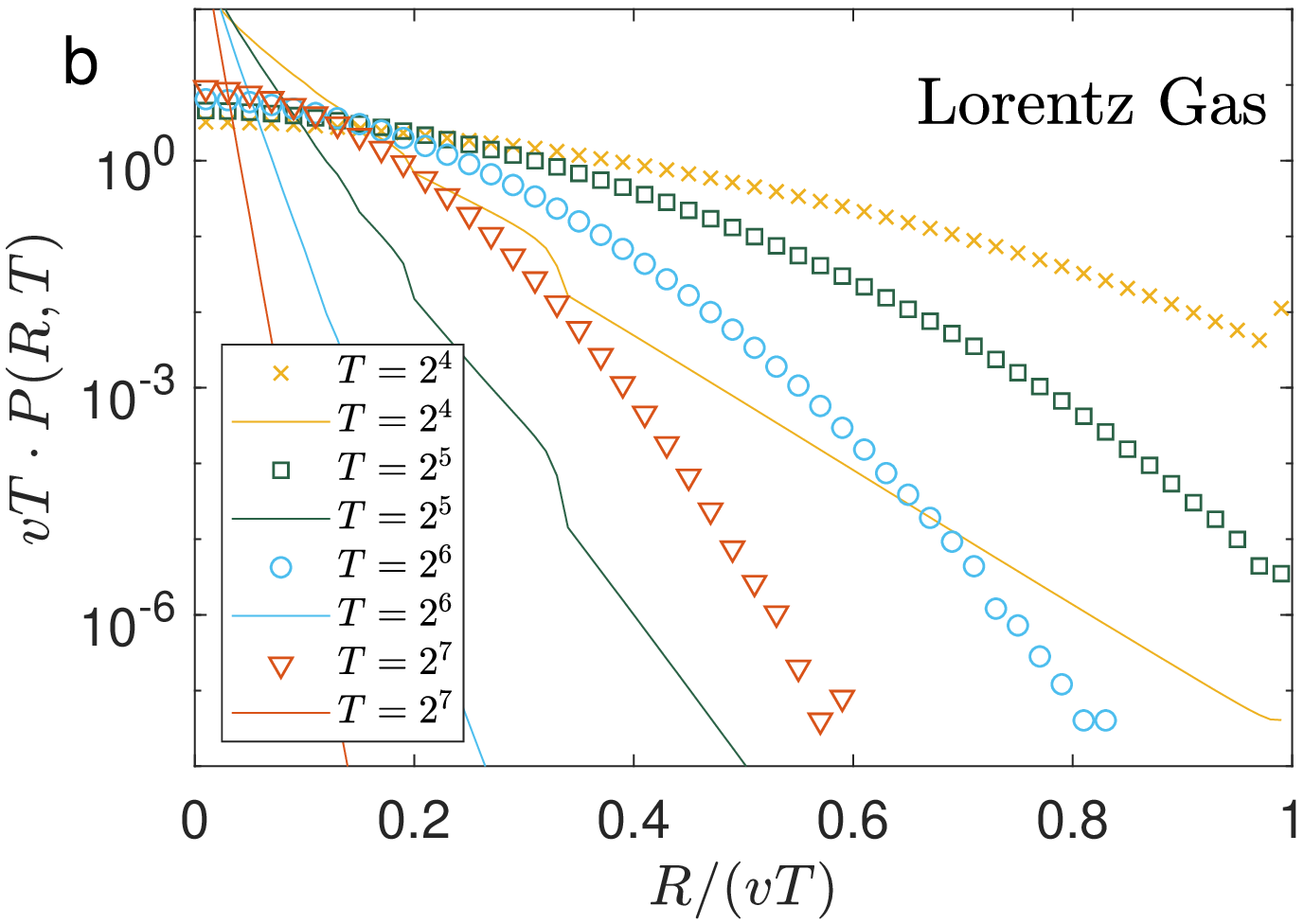}
		\includegraphics[width=0.49\textwidth]{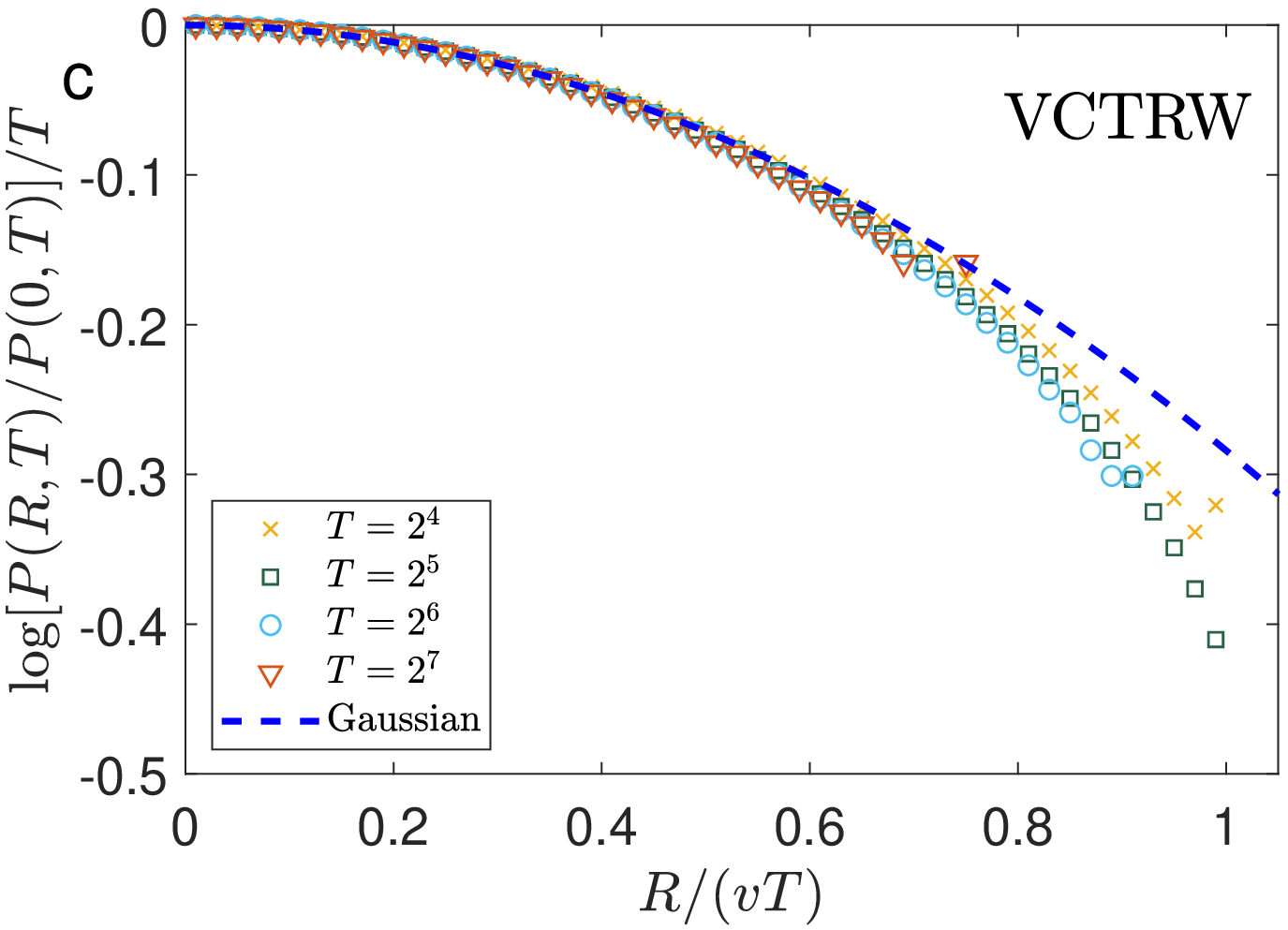}
	\includegraphics[width=0.49\textwidth]{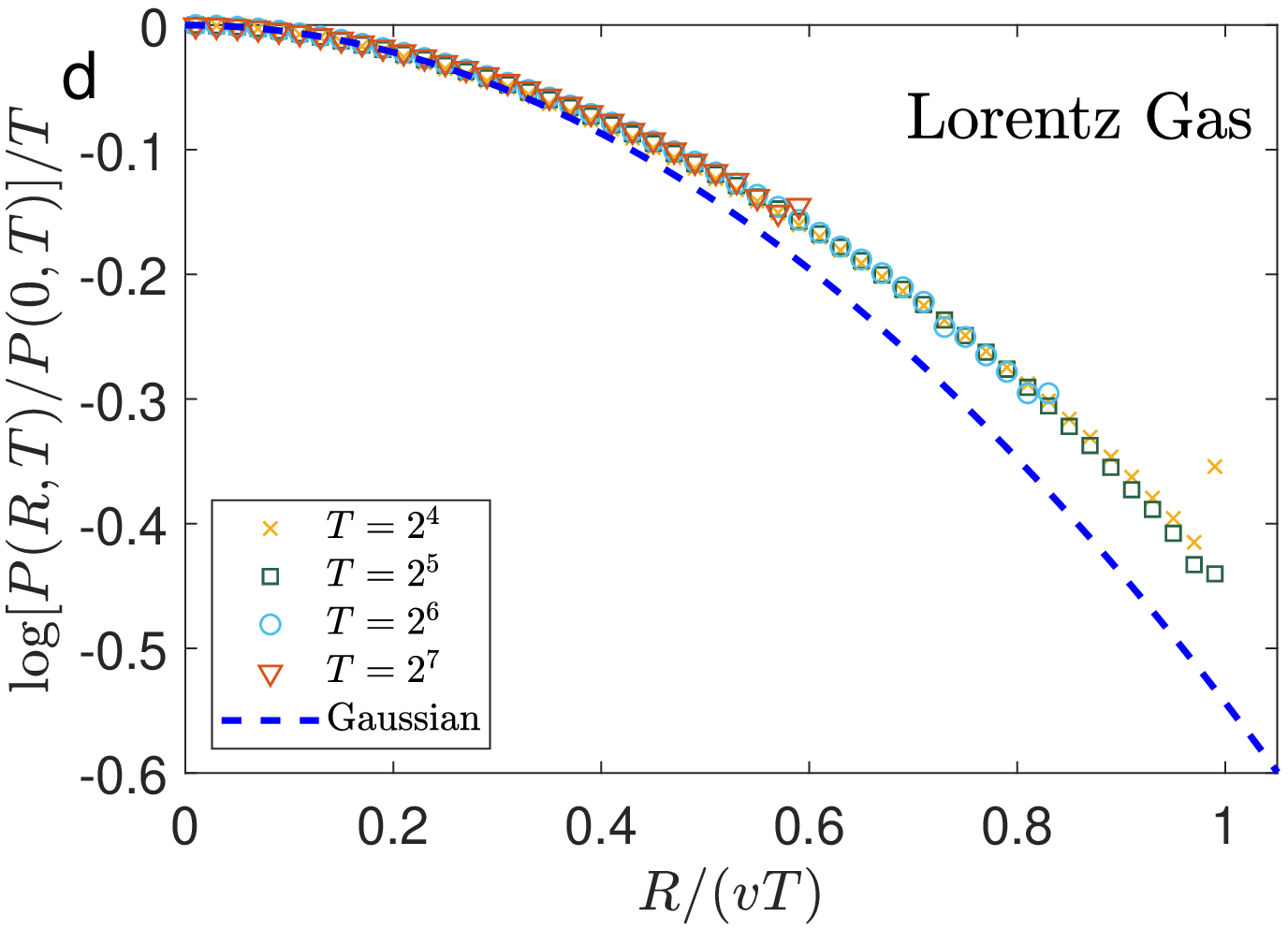}
	\caption{ The far tails of the distributions $P(R,T)$ for the L\'evy walk (panel a) and L\'evy Lorentz gas (panel b). We consider sharp distributions of times $t_i$ (panels a,c) and distances $L_i$ (panel b,d) using a Weibull distribution with $\alpha=1.1$. Moreover in both cases we fix $v=1$ with $\tilde t=1$ and $\tilde L=1$ respectively. The theoretical prediction in continuous lines are given by Eq. (\ref{BJ1}) in panel a and Eq. (\ref{single_j2}) in panel b. For  sharp distributions the big jump principle does not provide an estimate of the tails of the distributions since processes with multiple jumps cannot be neglected. In panels c and d we evidence that for $\alpha>1$ we can describe the system by using the standard large deviation approach. Dashed lines represent the Gaussian distributions that fits the PDF $P(R,T)$ at small $R$. }
	\label{weib1_5} 
\end{figure}

These properties imply that the big jump principle is an effective tool to describe the far tails of the distribution beyond the presence of  double scaling and strong anomalous diffusion, typical 
of  power law distributions. Figure \ref{weib1_5} also shows that, in the considered models, a sub-exponential decay is needed to apply the principle. 
The Weibull functions indeed describe also an exponential distribution for $\alpha=1$ or even a faster decay for $\alpha>1$. In Figure \ref{weib1_5} we compare the numerical simulation of the far tails of the distribution with the theoretical predictions in Eq.s (\ref{BJ1},\ref{single_j2}), both in the case of VCTRW (panel a) and Lorentz gas (panel b). The big jump prediction under-estimate the probability at large large $R$ by many order of magnitude and the difference asymptotically increases with time $T$. In this case, different processes involving many draws of the stochastic variable provide a contribution at large $R$, and they cannot be discarded.
On the other hand, for thin-tailed distributions with $\alpha>1$ panels c and d show that rare events can be described within the standard large deviation framework \cite{largedev} where
\begin{equation}
P(R,T)\sim e^{-T \cdot I(R/T)}
\label{dev}
\end{equation}  
asymptotically for large $T$. The unknown convex function $I(x)$ is called the {\em rate function}. An expansion around the minimum of $I(x)$ gives the Gaussian central limit theorem, but $I(x)$ also describes the non Gaussian tails of $P(R,T)$. The analytic expressions for $B(R,T)$ obtained with the big jump principle at $\alpha<1$ cannot be fitted into formula (\ref{dev}), so the big jump principle and the usual large deviation result appear as two complementary approaches, valid for super-exponential and sub-exponential distributions respectively.

\section{Conclusions}

In this paper we extend the validity of the big jump approach to the case of correlated random variable with stretched exponential distributions.
In particular, after critically reviewing the heuristic approach at the basis of the big jump calculations,  we apply to Weibull distributions the techniques introduced in \cite{VBB19} and we obtain the asymptotic functional form for the PDF of rare events both for VCTRW and for the Lorentz gas. The predicted function displays non-trivial analytic behaviors, encoding the non-universal nature of rare fluctuations. We check our results againt extensive numerical simulations which also show that the approach fails for distributions decaying faster than an exponential (i.e. the case $\alpha>1$). In that case, the usual large deviation approach in terms of rare function applies. Interestingly,  non-Gaussianity of rare events has been  recently considered in models with stretched exponential distributions for 
the driven run-and-tumble model \cite{gradenigo}.

Stretched exponential distributions feature  many differences with respect the power law case. The PDF does not display a single dynamical scaling length, strong anomalous diffusion is not present and the asymptotic scaling function of rare events is not an infinite density, so that techniques based on the resummation of anomalous moments cannot be applied \cite{Eli1,Eli2}. 
However, the big jump principle, which is based on the physical description of the process, turns out to be effective as well, providing the correct estimate of rare events. 
This opens new perspectives on a more rigorous approach to rare fluctuations in heavy tailed distributions.


\section*{References}


\begin{thebibliography}{}


\bibitem{Gumbel} Gumbel E J 2004 {\em Statistics of extremes} (Dover Publications,
Mineola)  
	
\bibitem{Hollander} 
den Hollander F 2008 {\em Large Deviations} (American Mathematical Society)
	
	
\bibitem{Holger}
Albeverio S, Jentsch V and Kantz H 2005
{\em Extreme Events in Nature and Society} 
(Springer, Berlin) 

\bibitem{Vulp} Vulpiani A, Cecconi F, Cancini M, Puglisi A and
Vergni D 2014 {\em Large Deviations in Physics: The legacy of the Law of Large Numbers} (Lecture Notes in Physics 995 Springer).


\bibitem{Embrechts} 
Embrechts P., Kappelberg C. and Mikosch T (1997) {\em Modelling Extremal Events for Insurance and Finance} (Springer)


 \bibitem{Lucilla1}
de Arcangelis L, Godano C, Grasso J R,  and Lippiello E 2016
Statistical physics approach to earthquake occurrence and forecasting
{\em  Phys. Rep.} \textbf{628}  1 


\bibitem{Chistyakov} 
Chistyakov V P 1964 A Theorem on Sums of Independent Positive Random Variables and Its Applications to Branching Random Processes
{\em  Theory of Probab. Appl.} \textbf{9} 640 


\bibitem{Foss} Foss S, Korshunov D and Zachary S 2013 {\em An introduction to heavy tailed and subexponential distributions} (Springer) 



\bibitem{Geluk} 
J. Geluk, Q. Tang {\em Asymptotic Tail Probabilities of Sums of Dependent
Subexponential Random Variables},
J. Theor. Probab.  \textbf{22} 871 (2009).


\bibitem{Clusel1} E. Bertin, and M. Clusel, 
{\em Generalized extreme value statistics and sum of correlated variables},
J. Phys. A.: Math. Theor. {\bf 39}, 7607 (2006).


\bibitem{Maj2} Szavits-Nossan J, Evans M R and Majumdar S N 2014 
Constraint-Driven Condensation in Large Fluctuations of Linear Statistics
\textit{ Phys. Rev. Lett.} \textbf{112}  020602

\bibitem{filias} Filiasi M,  Livan G, Marsili M,  Peressi M, Vesselli E and Zarinelli E 2014 
On the concentration of large deviations for fat tailed distributions, with application to financial data \textit{ J. Stat. Mech.: Theor. Exp.}  P09030 

\bibitem{corberi} Corberi F 2017
Development and regression of a large fluctuation
\textit{ Phys. Rev. } E \textbf{95} 032136 

\bibitem{WVBB19}
Wang W, Vezzani A, Burioni R and Barkai E 2019
Transport in disordered systems: the single big jump approach
arXiv:1906.04249


\bibitem{VBB19}
Vezzani A, Barkai E and Burioni R 2019
Single-big-jump principle in physical modeling
\textit{ Phys. Rev. } E \textbf{100} 012108 (2019)

\bibitem{zaburdaev}
Zaburdaev V, Denisov S and Klafter J 2015
L\'evy walks
{\em Rev. Mod. Phys.} \textbf{87}, 483 



\bibitem{VBB19b}
Microscopic dynamics in rare events: generalized L\'evy processes and the big jump principle
Vezzani A, Barkai E and Burioni R 2019
arXiv:1908.10975

\bibitem{Albers}
T. Albers, G. Radons, 
\textit{Exact Results for the Nonergodicity of d-Dimensional Generalized Lévy Walks},
Phys. Rev. Lett. \textbf{120} 104501
(2018)

\bibitem{Sokolov}
M. Bothe, F. Sagues, I.M. Sokolov,
\textit{Mean squared displacement in a generalized L\'evy walk model},
Phys. Rev. E \textbf{100} 012117 (2019)



\bibitem{levyrand}
Burioni R, Caniparoli L and Vezzani A 2010
{\em L\'evy walks and scaling in quenched disordered media},
{\em Phys. Rev. E} \textbf{81}, 060101(R) 


\bibitem{Fleurov}
Barkai E, Fleurov V and Klafter J 2000
One-dimensional stochastic L\'evy-Lorentz gas
{\em Phys. Rev.} E \textbf{61} 1164

\bibitem{castiglione}
Castiglione P, Mazzino A, Muratore-Ginanneschi P and Vulpiani A 1999
On strong anomalous diffusion
\textit{Physica } D \textbf{134} 75 


\bibitem{ctrw}
Montroll E W and Weiss G H 1965 
Random Walks on Lattices. II 
\textit{ J. Math. Phys.} \textbf{6} 167

\bibitem{sornette}
Laherr\'ere J and Sornette D 1998
Stretched exponential distributions in nature and economy: "fat tails" with characteristic scales
{\em Eur. Phys. J.} B \textbf{2} 525 
 

\bibitem{weibull}
Weibull W 1951
A Statistical Distribution Function of Wide Applicability
{\em J. Appl. Mech.-Trans.} ASME, \textbf{18} 293



\bibitem{Ub}
Burioni R, Ubaldi E and Vezzani A 2014
Superdiffusion and transport in two-dimensional systems with Lévy-like quenched disorder
\textit{ Phys. Rev.} E \textbf{89} 022135 

\bibitem{Bianchi}
Bianchi A, Cristadoro G, Lenci M and Ligab\`{o} M 2016
Random Walks in a One-Dimensional L\'evy Random Environment
\textit{ J. Stat. Phys.} \textbf{163}  22 

\bibitem{Magdziarza}
Magdziarz M and Szczotka W 2018
Diffusion limit of L\'evy-Lorentz gas is Brownian motion
\textit{ Commun. Nonlinear. Sci. Numer. Simul. } \textbf{69} 100 

\bibitem{Bianchi2}
Bianchi A, Lenci M and Pène F 2020
{Continuous-time random walk between L\'evy-spaced targets in the real line}
\textit{Stochastic Process. Appl.} \textbf{130} 708


\bibitem{Artuso}
Artuso R, Cristadoro G, Onofri M and Radice M 2018
Non-homogeneous persistent random walks and Lévy–Lorentz gas
\textit{ J. Stat. Mech.}  083209 

\bibitem{Eli1}
Rebenshtok A, Denisov S, Hanggi P and Barkai E 2014
Non-Normalizable Densities in Strong Anomalous Diffusion: Beyond the Central Limit Theorem
{\em Phys. Rev. Lett.} \textbf{112} 110601 

\bibitem{Eli2}
Rebenshtok R, Denisov S, Hanggi P and Barkai E 2014
Infinite densities for Lévy walks
{\em Phys. Rev.} E \textbf{90}  062135 

\bibitem{largedev}
Touchette H 2009
The large deviation approach to statistical mechanics
{\em  Phys. Rep.} \textbf{478}  1 


\bibitem{gradenigo}
Gradenigo G and Majumdar S N 2019
A first-order dynamical transition in the displacement distribution of a driven run-and-tumble particle
{\em  J. Stat. Mech. } 053206 


\end{thebibliography}
\end{document}